\documentclass[review]{elsarticle}

\usepackage{graphicx}
\usepackage{epstopdf}
\usepackage{subcaption}
\usepackage{color}
\usepackage{rotating}

\journal{Journal of \LaTeX\ Templates}

\begin{document}

\begin{frontmatter}



\title{Slow modes in the Hermean magnetosphere: effect of the solar wind hydrodynamic parameters and IMF orientation.}


\author[label1,label2]{J. Varela}
\ead{deviriam@gmail.com (telf: 0033782822476)}

\author[label3]{F. Pantellini}
\author[label3]{M. Moncuquet}

\address[label1]{AIM, CEA/CNRS/University of Paris 7, CEA-Saclay, 91191 Gif-sur-Yvette, France}
\address[label2]{LIMSI, CNRS-UPR 3251, Rue John von Neumann, 91405 Orsay, France}
\address[label3]{LESIA, Observatoire de Paris, CNRS, UPMC, Universite Paris-Diderot, 5
place Jules Janssen, 92195 Meudon, France}

\begin{abstract}

The aim of this study is to simulate the slow mode structures in the Hermean magnetosphere. We use a single fluid MHD model and a multipolar expansion of the Northward displaced Hermean magnetic field, to perform simulations with different solar wind parameter to foreseen the most favorable configuration for the formation of slow modes, attending to the solar wind density, velocity, temperature and the interplanetary magnetic field orientation. If the interplanetary magnetic field is aligned with the Mercury-Sun direction, the magnetic axis of Mercury in the Northward direction or the planet orbital plane, slow mode structures are observed nearby the South pole. If the orientation is in the Sun-Mercury or Northward directions, slow mode structures are observed nearby the North pole, but smaller compared with the structures near the South pole. Increase the density or the solar wind velocity avoids the formation of slow modes structures, not observed for a dynamic pressure larger than $6.25 \cdot 10^{-9}$ Pa in the case of a Northward interplanetary magnetic field orientation, due to the enhancement of the bow shock compression. If the solar wind temperature increases, the slow mode structures are wider because the sonic Mach number is smaller and the bow shock is less compressed. 

\end{abstract}


\end{frontmatter}


\section{Introduction}
\label{Introduction}

MESSENGER spacecraft observations revealed several characteristics of the Hermean magnetosphere as a northward shift of its dipolar field by $0.2$ of the planetary radius $(R_{M})$, a dipolar moment of $195$nT$*R^{3}_{M}$ and a tilt of the magnetic axis relative to the planetary spin axis smaller than $0.8^{0}$ \cite{2011Sci...333.1859A}. With the data of more than thousand of orbits in the North hemisphere, the Hermean magnetic field can be modeled by an axisymmetric multipolar expansion \cite{2012JGRA..11710228R,2012JGRE..117.0L12A}.

MESSENGER observations show a variable Herman magnetosphere due to the wide range of possible configurations of the solar wind (SW), leading to the formation of different magnnetospheric structures \cite{2009JGRA..11410101B,2011PandSS...59.2066B,2013JGRA..118...45B,2008Sci...321...82A,JGRE:JGRE3136}. Among these structures this study is focus in the slow modes. The slow modes are standing structures in the magnetosheath, extensively analyzed in the Earth magnetosphere \cite{1976JGR....81.1636Z,1992JGR....97.2873S,2004AnGeo..22.4273W} and recently in the Hermean magnetosphere \cite{2015PandSS..112....1P}. Numerical analysis of slow modes in Mercury predict the existence of slow modes fronts and eventually slow mode shocks just upstream of the magnetopause, particularly strong in the regions with large magnetic shear near the reconnection points. The presence of slow mode rarefaction fronts are also foretold forming standing structures in the magnetosheath as subproduct of the slow mode expansion \cite{PLA:96544}. The magnetic field lines in the compressional fronts turn connecting the IMF with the planetary magnetic field while in the rarefaction front the plasma flow and the magnetic field is diverted around the planet.

The aim of the present research is to simulate the interaction of the SW with the magnetic field of Mercury and the formation of slow modes in the Hermean magnetosphere for different configurations of the SW. First we fix the solar wind density, velocity and temperature to study the effect of the IMF orientation (for IMF module of $30$ nT): in the Mercury-Sun (Bx) and Sun-Mercury (Bxneg) directions, aligned with the Hermean magnetic field in the Northward (Bz) and Southward (Bzneg) direction, as well as in the planet orbital plane perpendicular to the previous directions oriented to the East (By) and to the West (Byneg). In the second part we study the effect of the hydrodynamic parameters (density, velocity and temperature) for a fixed IMF orientation (a Northward IMF orientation with module $30$ nT).

We use the MHD version of the single fluid code PLUTO in the ideal and inviscid limit for spherical 3D coordinates \cite{2007ApJS..170..228M}. The Northward displacement of the Hermean magnetic field is represented by a multipolar expansion \cite{2012JGRE..117.0L12A}. 

This paper is structured as follows. Section II, model description. Section III, effect of the IMF orientation. Section IV, effect of the hydrodynamic parameters. Section V, conclusion.   

\section{Numerical model}
\label{Model}

We use the MHD version of the code PLUTO in the ideal and inviscid limit for a single polytrophic  fluid in 3D spherical coordinates. The code is freely available online \cite{2007ApJS..170..228M}.

The simulation domain is confined within two spherical shells, representing the inner (planet) and outer (solar wind) boundaries of the system. Between the inner shell and the planet surface (at radius unity in the domain) there is a "soft coupling region" where special conditions apply (defined in the next section).The shells are at $0.6 R_{M}$ and $12 R_{M}$ ($R_{M}$ is the Mercury radius).

The conservative form of the equations are integrated using a Harten, Lax, Van Leer approximate Riemann solver (hll) associated with a diffusive limiter (minmod). The divergence of the magnetic field is ensured by a mixed hyperbolic/parabolic divergence cleaning technique (DIV CLEANING) \cite{2002JCoPh.175..645D}.

The grid points are $196$ radial points, $48$ in the polar angle $\theta$ and $96$ in the azimuthal angle $\phi$ (the grid poles correspond to the magnetic poles).

The planetary magnetic field is an axisymmetric model with the magnetic potential $\Psi$ expanded in dipolar, quadrupolar, octupolar and 16-polar terms \cite{2012JGRE..117.0L12A}:

$$ \Psi (r,\theta) = R_{M}\sum^{4}_{l=1} (\frac{R_{M}}{r})^{l+1} g_{l0} P_{l}(cos\theta) $$  
The current free magnetic field is $B_{M} = -\nabla \Psi $. $r$ is the distance to the planet center and $\theta$ the polar angle. The Legendre polynomials of the magnetic potential are:

$$ P_{1}(x) = x $$
$$ P_{2}(x) = \frac{1}{2} (3x^2 - 1) $$
$$ P_{3}(x) = \frac{1}{2} (5x^3 - 3x) $$
$$ P_{4}(x) = \frac{1}{2} (35x^4 - 30x^2 + 3) $$
the numerical coefficients $g_{l0}$ taken from Anderson et al. 2012 are summarized in the Table 1.

\begin{table}[h]
\centering
\begin{tabular}{c | c c c c}
coeff & $g_{01}$(nT) & $g_{02}/g_{01}$ & $g_{03}/g_{01}$ & $g_{04}/g_{01}$  \\ \hline
 & $-182$ & $0.4096$ & $0.1265$ & $0.0301$ \\
\end{tabular}
\caption{Multipolar coefficients $g_{l0}$ for Mercury's internal field.}
\end{table}

The simulation frame is such that the z-axis is given by the planetary magnetic axis pointing to the magnetic North pole and the Sun is located in the XZ plane with $x_{sun} > 0$. The y-axis completes the right-handed system

\subsection{Boundary conditions and initial conditions}

The outer boundary is divided in two regions, the upstream part (left in the figure) where the solar wind parameters are fixed and the downstream part (right in the figure) where we consider the null derivative condition $\frac{\partial}{\partial r} = 0$ for all fields. In the inner boundary the value of the intrinsic magnetic field of Mercury and the density are fixed. In the soft coupling region the velocity is smoothly reduced to zero in the inner boundary, the magnetic field and the velocity are parallel, and the profiles of the density is adjusted to keep the Alfven velocity constant $c_{a} =  B / (\mu_{0}\rho)^{1/2} = 25$ km/s with $\mu_{0}$ the magnetic permeability of the vacuum and $\rho$ the mass density. In the initial conditions we define a cone in the night side of the planet with zero velocity and low density centered in the planet. The IMF is cut off at $2 R_{M}$. We analyze the slow modes structures after the model evolution reaches the steady state. SW properties are kept constant during the simulation and the study doesn't describe any dynamic event in the Hermean magnetosphere. The transition from the initial conditions to the steady state only shows the numerical adjustment of the model without any physical valuable information, so it is not included in the text.

\section{IMF orientation}
\label{IMF orientation}

To study the effect of the IMF orientation in the slow mode structures we perform six simulations where the hydrodynamic parameters of the solar wind and the module of the IMF are fixed, changing only the orientation of the IMF in the directions described in the introduction. We include a reference case without IMF. We assume a fully ionized proton electron plasma, the sound speed is defined as $c_{s} = \sqrt {\gamma p / \rho}$ (with $p$ the total electron and proton pressure and $\rho = nm_{p}$ the mass density, $n$ the particle number and $m_{p}$ the proton mass), the sonic Mach number $M_{s} = v/c_{s}$ with $v$ the velocity. The solar wind velocity is aligned with the Sun-Mercury direction for simplicity in all the simulations. The fixed parameters in the simulations are summarized in the Table II. The configuration represents the expected properties of the solar wind during a phase of weak activity of the Sun. We choose this configuration with low dynamic pressure and temperature to maximize the effects driven by the IMF orientation in the slow mode structure.

\begin{table}[h]
\centering
\begin{tabular}{c | c c c c c c c c}
n (cm$^{-3}$) & $T$ K & $V$ (km/s) & $M_{s}$ & IMF (nT)  \\ \hline
$60$ & $58000$ & $250$ & $6.25$ & $30$\\
\end{tabular}
\caption{Fixed parameters in the simulations.}
\end{table}

The Fig.1 show the pressure distribution in a polar cut for the different orientation of the IMF. The first test to localize a slow mode structure is to search in the inner magnetosphere for a second maximum of the pressure (necessary but not sufficient condition). No second maximum is observed in the reference case and (G) the Bzneg orientation (H). There is a second maximum of the pressure nearby the North pole for the orientations Bxneg (B), By (C) and Bz (E) and nearby the South pole for the orientations Bx (A) and Bz (F). The Bz orientation shows the widest regions over both poles. The Bx and Bxneg orientations as well as the By and Byneg orientations show a more localized second maximum observed on the planet day side. The secondary maximum in the By and Byneg orientations is detached from the bow shock while for the Bx and Bxneg orientation it is attached. The red line in the plots defines the region where $v/v_{\phi} \leq 1$ with:

$$ v_{\phi}^{2} = \frac{1}{2} ({c_{s}^{2} + c_{a}^{2}} \pm [(c_{s}^{2} + c_{a}^{2})^{2} + 4c_{s}^{2}  c_{a}^{2} sin^{2}\theta_{kB}]^{1/2}) $$ 
the phase velocity of the mode with wave vector $\vec{k}$, defined as the normal vector to the slow/fast mode structure, and $\theta_{kB}$ the angle between the magnetic field and the mode wave vector. The phase velocity of the slow (fast) mode is smaller (larger) than the sound speed. One condition for the formation of the slow modes is that $v/v_{\phi} \leq 1$. The Fig. 1 shows that the secondary maximum of the pressure in the inner magnetosphere are fully or partially located inside a region with $v/v_{\phi} \leq 1$. The pink lines from the bow shock upstream to the inner magnetosphere indicate the region plotted in the graphs of the Fig. 2.

\begin{figure}[h]
\centering
\includegraphics[width=0.8\textwidth]{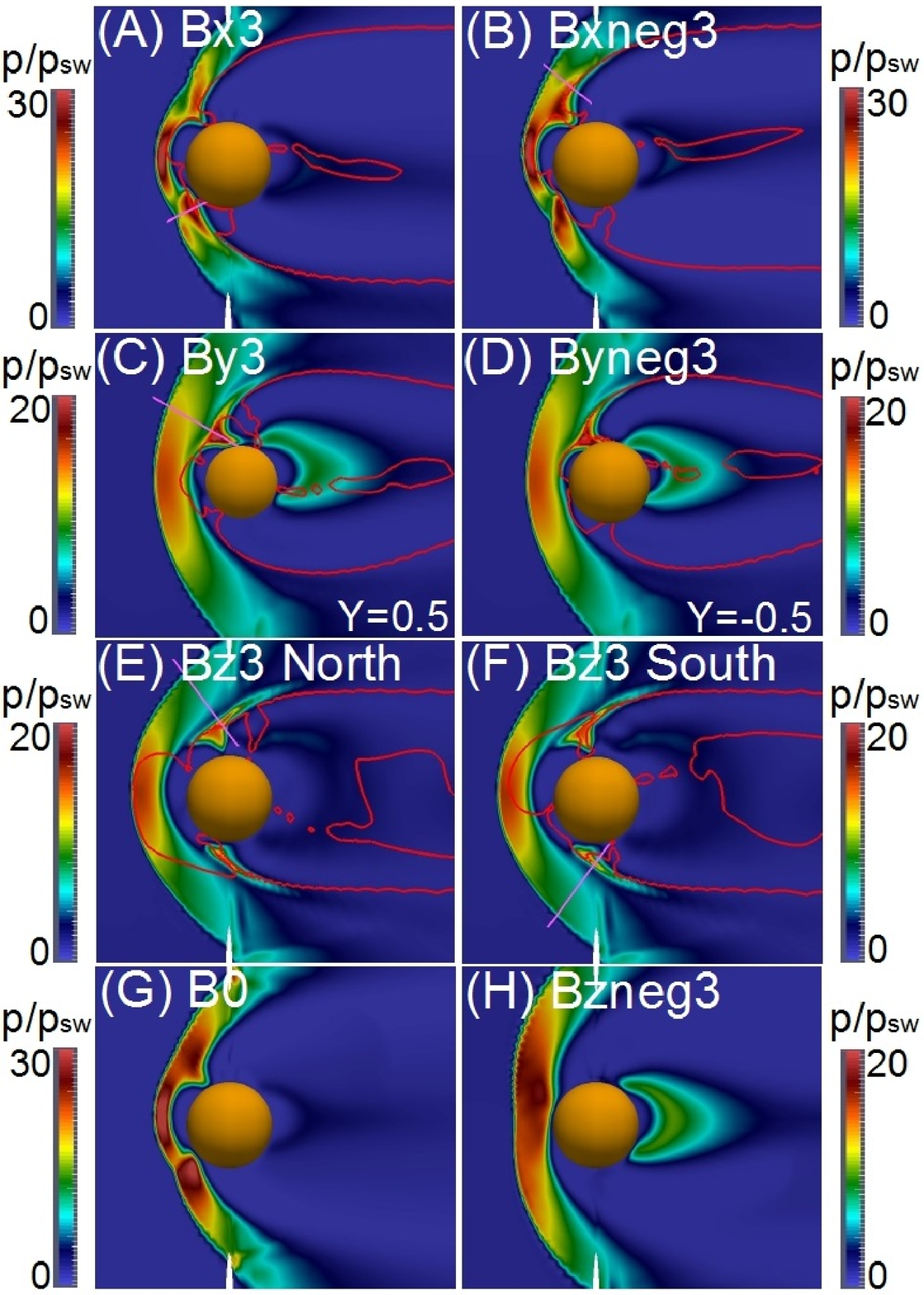}
\caption{Polar cut of the pressure distribution normalized to the solar wind pressure for the IMF orientations Bx (A), Bxneg (B), By (C), Byneg (D), Bz North pole (E), Bz South pole (F), reference case (G) and Bzneg orientation (H). The red line shows the region where $v/v_{\phi} \leq 1$. The profiles in the figure 2 are calculated along the solid pink lines. The plots are displaced 0.1 $R_{M}$ in Y direction except the By and Byneg cases that are displaced $\pm$0.5 $R_{M}$ in Y direction.}
\end{figure}

The Fig. 2 shows plots of the density, pressure, module of the velocity and magnetic field as well as the angle subtended between the velocity and magnetic fields $\theta_{vB}$ from the bow shock upstream to the inner magnetosphere, perpendicular to the slow mode front. We include in the plots the parallel compressibility defined as:
$$C_{p} = \frac{|\delta n|}{n} \frac{B}{|\delta B_{||}|}$$
with $|\delta n| = [(\partial n/ \partial x)^{2} + (\partial n/ \partial y)^{2} + (\partial n/ \partial z)^{2}]^{1/2}$ the variation of the number density and $|\delta B_{||}| = |\delta \vec{B} \cdot \vec{k}| $ the variation of the magnetic field intensity in the direction of the mode wave vector. If $C_{p} < 0$ ($C_{p} > 0$) it is a slow (fast) mode. There are two different regions in the graphs, the rarefaction and the compressional fronts of the standing structure (the compressional front is indicated by a black arrow). The compressional front is identified as a sharp increase of the pressure and a local maximum of the density in anti phase with the velocity and magnetic field modules. The rarefaction front is located between the BS and the compressional fronts where there is a slightly drop of the density and pressure, and a slightly increase of the velocity and magnetic field modules. For the Bx orientation the compressional front is sharper than for the Bxneg case but $C_{p}$ is not negative, almost null nearby the compressional front. $C_{p}$ is negative for the Bxneg orientation but the local minimum is slightly decorrelated with the compressional front. For the By orientation the compressional front is weaker, there is only a small drop of the magnetic field module, and a negative $C_{p}$ slightly decorrelated with the compressional front.  The Bz orientation show a sharp maximum of the pressure and minimum of the magnetic and velocity field in the compressional front at both poles. There is a region with $C_{p} < 0$ in the South pole located  around the compressional front, while in the North pole it is almost null in the compressional front but no negative. For all the orientations the angle $\theta_{vB}$ shows that the magnetic and velocity fields are perpendicular in the compressional front. 

\begin{figure}[h]
\centering
\includegraphics[width=0.8\textwidth]{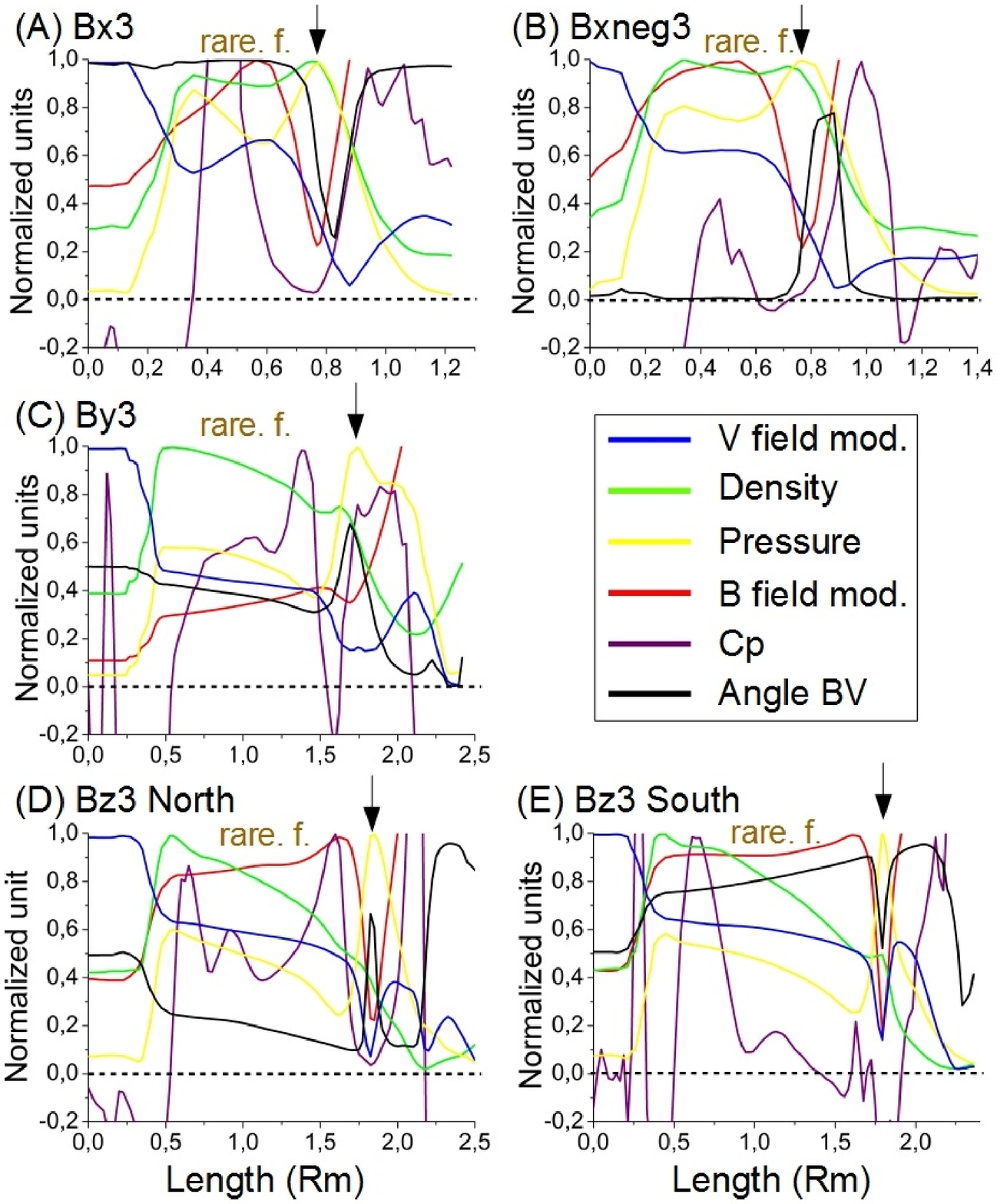}
\caption{Normalized module of the magnetic and velocity fields, density, pressure, the $C_{p}$ parameter and the angle between the velocity and the magnetic fields $\theta_{vB}$ along the lines plotted in the figure 1 crossing the slow modes structures for the Bx (A), Bxneg (B), By (C), Bz North pole (D) and Bz South pole (E) orientations.}
\end{figure}

The IMF orientation that leads to the formation of the widest slow modes structures, with clear patterns of the presence of compressional and rarefaction fronts in regions with $C_{p} < 0$, is the Northward orientation, particularly nearby the South pole. The optimal configuration to observe slow modes nearby the North pole is a combination of a Northward and Sun-Mercury IMF orientation, because for a pure Northward orientation the rarefaction front is not located in a region with $C_{p} < 0$ while for a pure Sun-Mercury orientation the compressional front is weaker, without a sharp local maximum of the pressure.

\section{Hydrodynamic parameters}
\label{Hydrodynamic parameters}
 
In this section we analyze the effect of the hydrodynamic parameters of the SW in the slow modes structure: density, velocity and temperature. We perform new simulations fixing a Northward IMF orientation, because it is the case with widest slow modes structures as we observed in the previous section.

The Fig. 3 shows the effect of the SW density in the slow mode structure for simulations with half, 2 times and 3 times the particle number of the reference case. If the SW density increases the region with $C_{p} < 0$ correlated with the compressional front is smaller in the South pole, while in the North pole the $C_{p}$ value is larger and positive (for a lower value of the density compared with the reference case, the $C_{p}$ value is almost null in the compressional front). In both poles the local maximum of the pressure and the module of the magnetic field is less sharper as the density increases (compressional front is weaker). The isolines of the $v/v_{\phi} \leq 1$, red (orange) for the wave vector in the North (South) pole, shows that the slow mode is smaller as the density increases and it is more concentrated around the standing structure. These results point out that a more compressed BS avoids the formation of the slow modes.

\begin{figure}[h]
\centering
\includegraphics[width=1\textwidth]{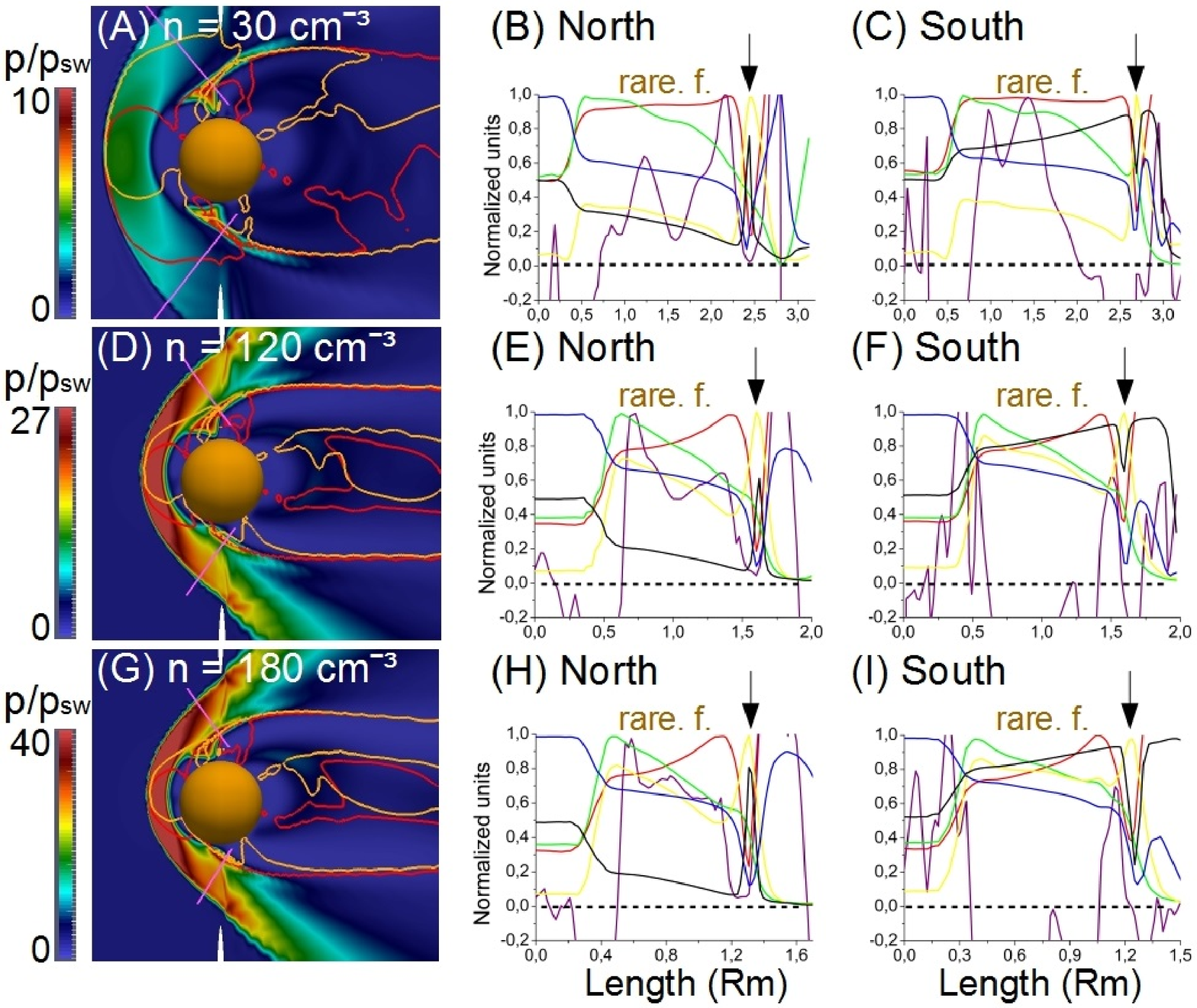}
\caption{Effect of the solar wind density in the slow mode structures. Polar cut of the pressure distribution normalized to the solar wind pressure for half (A), 2 times (D) and 3 times (G) the reference case. The red (orange) line shows the region where $v/v_{\phi} \leq 1$ for the wave vector in the North (South) pole. Pink lines show the plotted region crossing the slow modes structures: Module of the magnetic and velocity fields, density, pressure, $C_{p}$ parameter and angle between the velocity and the magnetic fields $\theta_{vB}$ for the simulation with half the density of the reference case, North pole (B) and South pole (C), 2 times, North pole (E) and South pole (F) and 3 times, North pole (H) and South pole (I). The plots are displaced 0.1 $R_{M}$ in Y direction.}
\end{figure}

The Fig. 4 shows the effect of the SW velocity in the slow mode structure for simulations with $200$, $350$ and $450$ km/s. The conclusions are the same than in the study of the density in both poles: for larger velocity the BS is more compressed and the local maximum (minimum) of the density and pressure (velocity and magnetic field modules) are less sharper, the regions with $C_{p} < 0$ decrease and the $v/v_{\phi} \leq 1$ isocontours are more located in the standing structures. For a slower SW the compressional front is sharper and the $C_{p} < 0$ region is bigger.
 
\begin{figure}[h]
\centering
\includegraphics[width=1.0\textwidth]{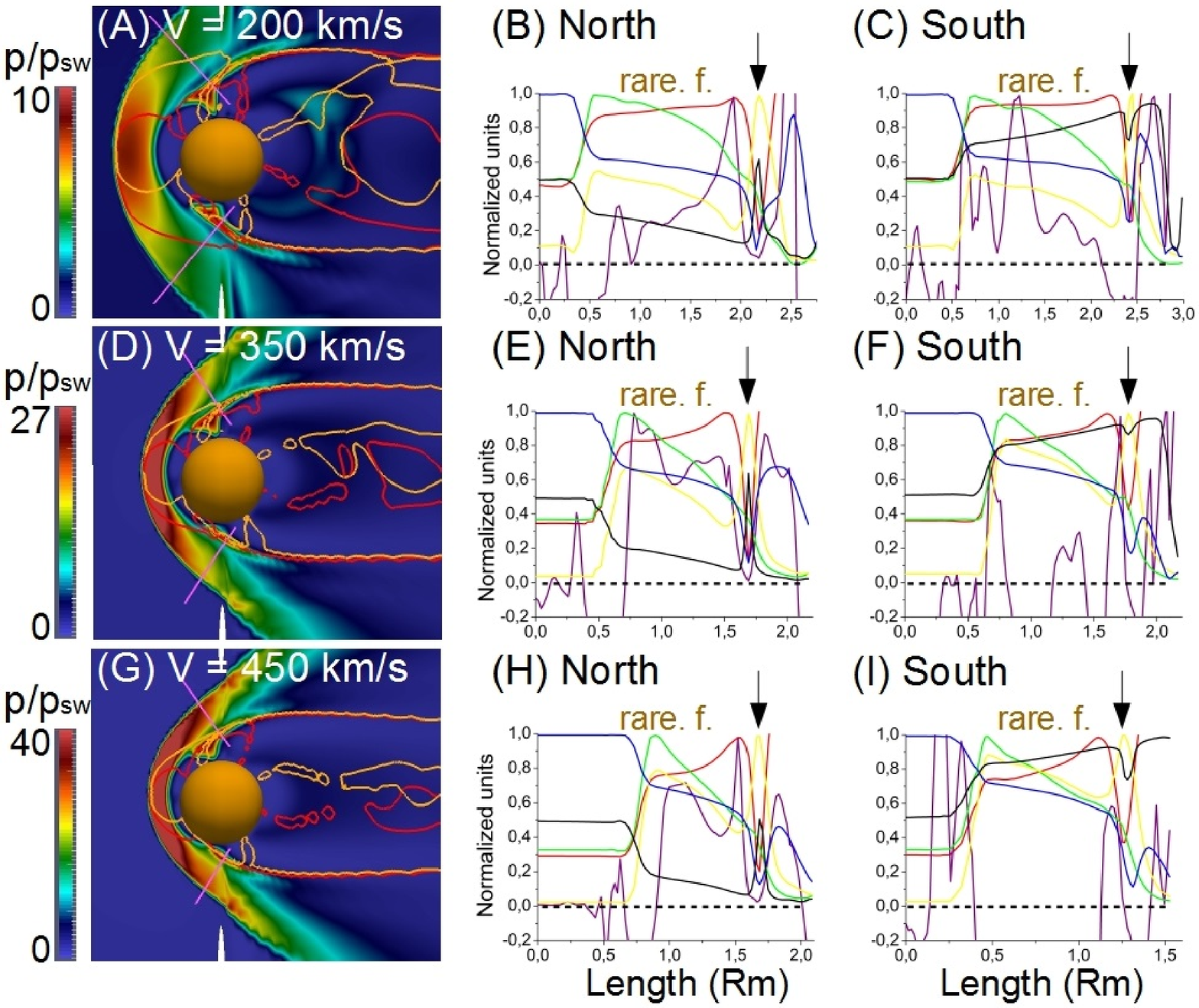}
\caption{Effect of the solar wind velocity in the slow mode structures. Polar cut of the pressure distribution normalized to the solar wind pressure for $200$ (A), $350$ (D) and $450$ (G) km/s. The red (orange) line shows the region where $v/v_{\phi} \leq 1$ for the wave vector in the North (South) pole. Pink lines show the plotted region crossing the slow modes structures: Module of the magnetic and velocity fields, density, pressure, $C_{p}$ parameter and angle between the velocity and the magnetic fields $\theta_{vB}$ for the simulation with $200$ km/s, North pole (B) and South pole (C), $350$ km/s, North pole (E) and South pole (F) and $450$ km/s, North pole (H) and South pole (I). The plots are displaced 0.1 $R_{M}$ in Y direction.}
\end{figure}

The Fig. 5 shows the effect of the SW temperature in the slow mode structure for simulations with $T = 100,000 $ and $T = 150,000 $ K. The structure of the compressional and rarefaction fronts are almost the same than in the reference case for both simulations, as well as the isocontours with $v/v_{\phi} \leq 1$, but the region with $C_{p} < 0$ (small $C_{p}$ values) in the South pole (North pole) is larger for hotter plasmas, and the angle between the velocity and magnetic field are closer to be perpendicular in the compressional front. This is consequence of the decompression of the BS when the SW temperature increases, because the sound velocity increases and the sonic Mach number drops. 

\begin{figure}[h]
\centering
\includegraphics[width=1.0\textwidth]{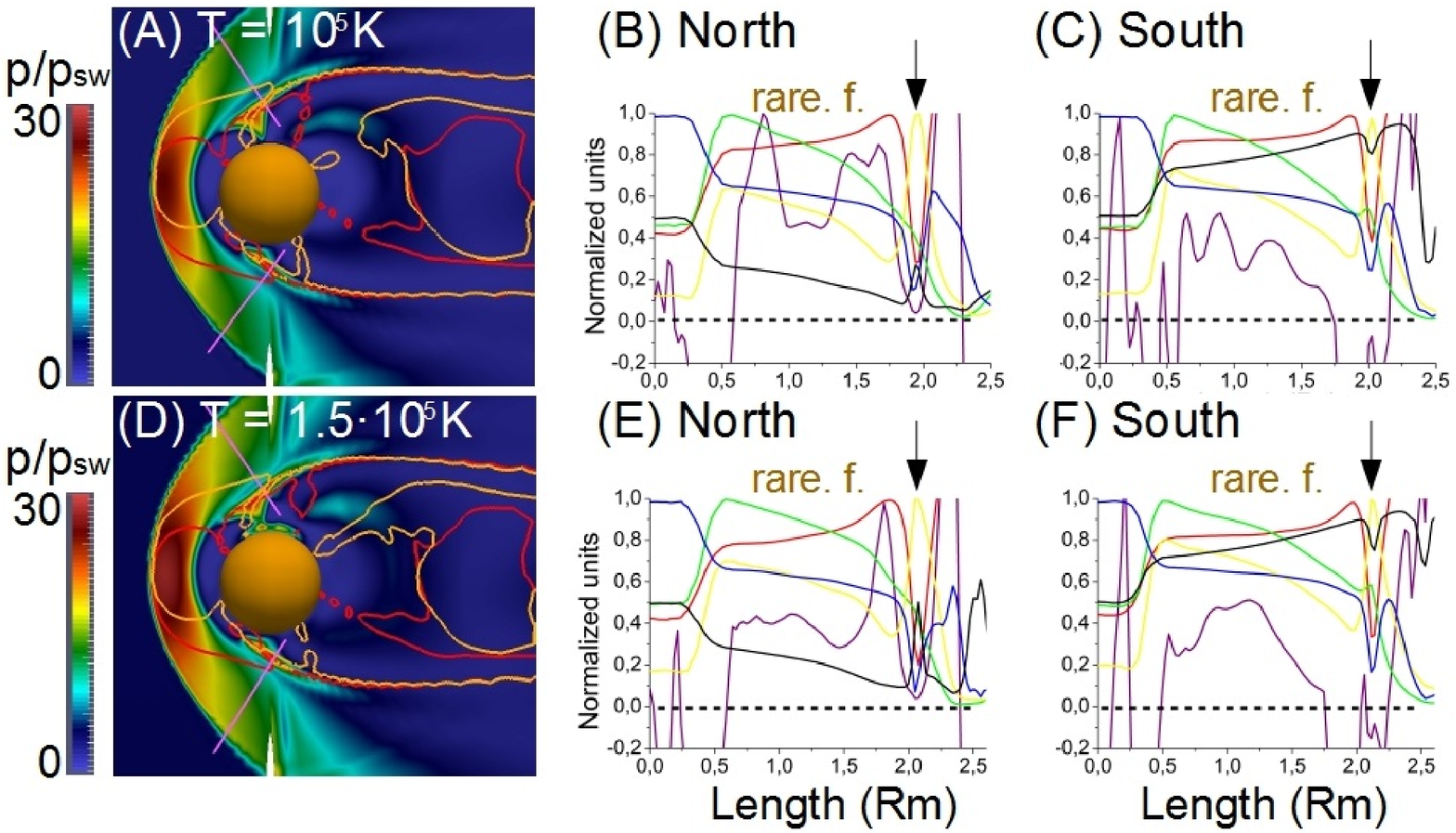}
\caption{Effect of the solar wind temperature in the slow mode structures. Polar cut of the pressure distribution normalized to the solar wind temperature of $T = 100,000 $K (A) and $T = 150,000$ K (D). The red (orange) line shows the region where $v/v_{\phi} \leq 1$ for the wave vector in the North (South) pole. Pink lines show the plotted region crossing the slow modes structures: Module of the magnetic and velocity fields, density, pressure, $C_{p}$ parameter and angle between the velocity and the magnetic fields $\theta_{vB}$ for the simulations with $T = 100,000$ K, North pole (B) and South pole (C) and $T = 150,000$ K, North pole (E) and South pole (F). The plots are displaced 0.1 $R_{M}$ in Y direction.}
\end{figure}

According to these results, if the SW dynamic pressure is $6.25 \cdot 10^{-9}$ Pa or larger, the slow modes are not observed in the Hermean magnetosphere even for the most favorable (Northward) IMF orientation. For a SW dynamic pressure of $3 \cdot 10^{-9}$ Pa or smaller, there are slow modes structures for a Northward IMF orientation. The enhancement of the BS compression leads to a drops of the magnetopause stand off distance, located too close to the planet and avoiding the formation of the slow modes that are advected by the SW flow or precipitate on the planet. The slow mode structure in the South pole is, for any configuration of the hydrodynamic parameters, wider than in the North pole. 

\section{Conclusions}
\label{Conclusions}

The IMF orientation and the hydrodynamic parameters of the SW modifies the properties of the slow modes structure in the Hermean magnetosphere. A Northward orientation is the optimal IMF orientation for the formation of slow modes in the South pole, but it is a combination of a Northward and Sun-Mercury orientation the optimal case in the North pole, leading to the sharpest local maximum (minimum) of the pressure and density (velocity and magnetic field modules) as well as the widest region with $C_{p} < 0$. 

The study of the hydrodynamic parameters indicate that a SW configuration with a large dynamic pressure, $6.25 \cdot 10^{-9}$ Pa or larger, avoids the formation of slow modes structures even for the optimal IMF orientation, but are observed for a dynamic pressure of $3 \cdot 10^{-9}$ Pa or smaller, pointing out that a large compression of the BS and the decrease of the magnetopause stand off distance lead to the advection of the slow mode structures by the solar wind or its precipitation on the planet surface. In the case of a SW configuration with a high temperature plasma, the BS compression drops leading to an extension of the region with $C_{p} < 0$. 

The limitation of the measurements of the Hermean magnetic field by MESSENGER to the North Hemisphere, added to the lack of information of the in situ SW density, velocity and temperature, make a difficult task to perform realistic simulations of the slow modes in the Hermean magnetosphere. In a future study we will select MESSENGER orbits with an optimal orientation of the IMF for the formation of slow modes nearby the North pole and to perform simulations in the cases where the SW model predicts density, velocity and temperature conditions on the range of values favorable for the formation of slow modes. Preliminary results show a correlation between local downfalls of the magnetic field measured by MESSENGER, with slow mode structures observed in simulations performed using similar conditions of solar wind and IMF orientation, for example during the satellite orbit of 2011 September 08 (IMF orientation dominated by the Northward component and dynamic pressure of $5.42 \cdot 10^{-9}$ Pa.)

\section{Aknowledgments}
The research leading to these results has received funding from the European Commission's Seventh Framework Programme (FP7/2007-2013) under the grant agreement SHOCK (project number 284515).

\section*{References}

\bibliography{mybibfile}

\end{document}